# Data-driven formulation of natural laws by recursive-LASSO-based symbolic regression

Yuma Iwasaki[1,2,3]* & Masahiko Ishida[1,3]

[1]System Platform Research Laboratories, NEC Corporation, Kawasaki 211-8666, Japan

[2]PRESTO, JST, Saitama 322-0012, Japan

[3]NEC-AIST Quantum Technology Cooperative Research Laboratory, National Institute of Advanced Industrial Science and Technology (AIST), Tsukuba, 305-8568, Japan

**Abstract**

Discovery of new natural laws has for a long time relied on the inspiration of some genius. Recently, however, machine learning technologies, which analyze big data without human prejudice and bias, are expected to find novel natural laws. Here we demonstrate that our proposed machine learning, recursive-LASSO-based symbolic (RLS) regression, enables data-driven formulation of natural laws from noisy data. The RLS regression recurrently repeats feature generation and feature selection, eventually constructing a data-driven model with highly nonlinear features. This data-driven formulation method is quite general and thus can discover new laws in various scientific fields.

**Introduction**

Development of natural law formulation, describing the world by mathematical expressions, has been based on data observation and consideration by scientists. For example, there is a story that Isaac Newton observed an apple falling from a tree and, after deep deliberation, had a brainstorm for the law of universal gravitation (There are also other versions of this story). Natural laws have been formulated by only a small number of geniuses.

Recently, machine learning modeling using big data has been expected to contribute to data-driven discovery of new natural laws because data on natural phenomena (e.g., physical, chemical, biological, and astronomical phenomena) are being accumulated rapidly all over the world[1-9]. Data-driven approaches by machine learning have the potential to excavate from such big data a novel rule that we have never noticed[10,11]. For instance, Baldi et al. used deep learning to construct a high-energy-physics model of particle colliders[12]. Their deep learning model was highly predictive, which means that data-driven model construction was somehow made successful by machine learning.

However, it is difficult for scientists to understand phenomena represented by deep learning models because the models are hard to interpret. A deep learning model is expressed as connections between large numbers of perceptrons (neurons), and its interpretation by humans is not feasible. Natural law is discovered only when humans recognize and understand a phenomenon modeled as a simple mathematical expression. The processes of machine learning modeling are in themselves insufficient for discovering new physical rules and laws.

For focusing on the modeling of nature laws, it is ideal to construct a data-driven model as a nonlinear mathematical expression. If machine learning automatically generates the nonlinear mathematical expression model from observed data without prior knowledge, it can directly discover a new natural law. Symbolic regression techniques using a genetic algorithm (GA) allow constructing a data-driven nonlinear mathematical expression model[13-21].

Here we propose a simple symbolic regression, recursive-LASSO-based symbolic regression method (referred to here as the RLS regression method), for constructing a data-driven nonlinear mathematical expression model and demonstrate that the proposed method can perform data-driven formulation of natural laws from noisy data. Figure 1 gives a rough explanation for the RLS regression, showing iteration of a nonlinear features generation step and a feature selection step. For simplicity of explanation, we here assume only three original features $x = \{x_1, x_2, x_3\}$. RLS regression firstly makes nonlinear features using the four basic arithmetic operations (and squaring) with the original features $x$. Figure 2a shows the generated nonlinear features. Secondly, feature selection (LASSO[22]) is performed with these features and the selected features are regarded as new features $x'$. Here we assume that three features $x' = \{x_2, x_1 x_2, x_2 + x_3\}$ are selected. Again, the RLS regression makes some nonlinear features by using both original features $x$ and the new features $x'$ with de-duplication (i.e., $x_2$). Figure 2b shows the secondary nonlinear features. Note that the features generated second (Fig. 2b) are more complex than those generated first (Fig. 2a). The RLS regression searches for proper nonlinear features automatically by repeating this cycle recursively. Details of this algorithm and parameter settings are provided in the Methods section.

## Results

As simple examples, we demonstrate that the RLS regression can represent two nonlinear natural phenomena. The parameter setting in the RLS regression, which is shown in the Methods section, is the same for both example cases. The data sets used are synthetic with Gaussian noise (standard deviation $\sigma$).

**Data-driven formulation of small particles falling.**

Figure 3a shows an illustration of a falling particle in a fluid. The particle radius $D_p$ and particle density $\rho_p$, are respectively 0.0001–0.0002 m and 5000–6000 kg/m$^3$ (values shown here as random value range). The fluid density $\rho_f$ is 1000–2000 kg/m$^3$ and fluid viscosity $\eta$ is 0.0008–0.012 Pa·s. The gravitational acceleration $g$ is assumed to be 9.8 m/s$^2$. The number of data sets (observations) is 1000. The terminal velocity $v_s$ m/s includes Gaussian noise with standard deviation $\sigma$ = 0.005. With the $D_p$, $\rho_p$, $\rho_f$, and $\eta$, we tried formulating the terminal velocity models $v_s = f(D_p, \rho_p, \rho_f, \eta)$ by the RLS regression method. For comparison, the LASSO and Neural Network (NN) models, which do not need a manual listing of potential constant values and potential nonlinear features beforehand, were also constructed.

Figures 3b, c, and d show the accuracy of the LASSO, NN, and RLS regression models, respectively. The four-fold cross-validation mean absolute errors (CV-MAEs) of the LASSO, NN, and RLS regression models are $3.82 \times 10^{-3}$, $2.95 \times 10^{-3}$, and $2.96 \times 10^{-3}$, respectively. Thus although as shown in Fig. 3e the LASSO makes an interpretable mathematical model, its CV-MAE is larger than that of the NN and RLS regression models. This is to be expected because this modeled phenomenon is not expressed by a simple linear combination of the $D_p$, $\rho_p$, $\rho_f$, and $\eta$. The NN model, on the other hand, has better prediction accuracy because the highly flexible model can express this nonlinear relationship. As shown by the inset in Fig. 3c, however, it is quite difficult for a person to interpret the complex NN model expressed as connections between large numbers of perceptrons (neurons). The prediction accuracy of the RLS regression model is comparable to that of the NN model. Moreover, we can understand the interpretable RLS regression model, which accordingly is of great help in understanding the physics of the modeled phenomenon.

When we ignore small regression coefficient terms, the RLS regression model becomes as follows.

$$v_s = -0.0039 + 0.602 \frac{D_p^2(\rho_p - \rho_f)}{\eta} - 0.00000147 \left(\rho_p - \frac{\rho_p^2}{D_p} + \rho_f\right)$$

$$v_s \approx 0.602 \frac{D_p^2(\rho_p - \rho_f)}{\eta} \tag{1}$$

This model is consistent with a known natural law, Stokes' law[23].

$$v_s = \frac{D_p^2(\rho_p - \rho_f)g}{18\eta} \tag{2}$$

Thus the RLS regression method can construct a data-driven nonlinear mathematical expression model describing Stokes' law without any prior knowledge. Although the NN model's high prediction accuracy suggests that it too represents Stokes' law, scientists cannot understand the inside of the model, so it does not inspire them to investigate the modeled phenomenon.

**Data-driven formulation of rigid-body collision.**

To demonstrate a case with a larger number of original features, we next model a rigid spherical body (ball) collision phenomenon as shown in Fig. 4a. Firstly, a red ball with mass $m_1$ (1–30 g) is rolling on the hill at a velocity $v_1$ (0.01–0.3 m/s). At the foot of a slope, a blue ball with mass $m_2$ (1–10 g) is rolling at a velocity $v_2$ (0.01–0.1 m/s). After a few seconds, the red ball accelerates on the slope and collides with the blue ball. Finally, the velocities of the red and blue balls become $v_1'$ (0.01–1 m/s) and $v_2'$ (0.01–3 m/s), respectively. The height of the hill is L, including Gaussian noise with $\sigma$ = 0.03. The number of data sets (observations) is 1000. The gravitational acceleration $g$ is assumed to be 9.8 m/s$^2$. A green ball with mass $m_3$ (1–50 g) just stays on the hill from beginning to end. With the $m_1$, $m_2$, $m_3$, $v_1$, $v_2$, $v_1'$, and $v_2'$, we tried formulating the data-driven model $L = f(m_1, m_2, m_3, v_1, v_2, v'_1, v'_2)$ by the RLS regression method. For comparison, the LASSO and NN models were also constructed.

Figures 4b, c, and d show the accuracies of the LASSO, NN, and RLS regression models. The LASSO and RLS regression model equations are shown in Fig. 4e. The LASSO correctly removed the $m_3$ term, which clearly has nothing to do with the collision phenomenon shown in Fig. 4a, by feature selection with $L_1$ regularization. However, the CV-MAE for the LASSO model is high because this phenomenon cannot be expressed by a simple linear combination of the $m_1$, $m_2$, $m_3$, $v_1$, $v_2$, $v_1'$, and $v_2'$. The NN and RLS regression models have better prediction accuracy because these highly flexible models can express nonlinear relationships between variables. The NN model, whose visualization is shown in Fig. 4c, has low interpretability, while the RLS regression model is highly interpretable and thus we can easily investigate the physics of the phenomenon.

Simplifying the RLS regression model by ignoring the small regression coefficient term and intercept term, it becomes as follows.

$$L = 0.01 - 0.058(v_1^2 - v'^2_1) - 0.059\frac{m_2}{m_1}v'_2(v_2 - v'_2) - 0.06\frac{m_2}{m_1}v_2(v_2 - v'_2)$$

$$L \approx 0.058(v'^2_1 - v_1^2) + 0.059\frac{m_2}{m_1}v'^2_2 - 0.06\frac{m_2}{m_1}v_2^2 \quad (3)$$

This model is consistent with a correct answer equation derived by applying the law of the conservation of energy, where the kinetic energy ($\frac{1}{2}mv^2$) and potential energy ($mgL$) are conserved before and after collision.

$$m_1 gL + \frac{1}{2}m_1 v_1^2 + \frac{1}{2}m_2 v_2^2 = \frac{1}{2}m_1 v'^2_1 + \frac{1}{2}m_2 v'^2_2$$

$$L = \frac{1}{2g}(v'^2_1 - v_1^2) + \frac{1}{2g}\frac{m_2}{m_1}v'^2_2 - \frac{1}{2g}\frac{m_2}{m_1}v_2^2 \quad (4)$$

We demonstrated that the RLS regression allows data-driven models to be constructed as interpretable mathematical expressions. Because we need to confirm that the RLS regression models are consistent with the physically true model, in this paper we show the physically known phenomena as examples. When the RLS regression method is applied to an unsettled issue (e.g., chaotic phenomenon, astronomical phenomenon, unknown compound), we can find a new rule expressed mathematically.

**Discussion**

Figures 5a and b show the prediction accuracies of the NN and RLS regression models for no-noise data ($\sigma$ = 0) in the case study of the rigid-body-collision modeling. The prediction accuracy of the RLS regression model was better than that of the NN model. This is to be expected because the RLS regression model succeeded in describing a physically true model. In this case, the RLS regression model was created as

$$L = -0.00000805 - 0.051(v_1^2 - v'^2_1) - 0.051\frac{m_2}{m_1}v'_2(v_2 - v'_2) - 0.051\frac{m_2}{m_1}v_2(v_2 - v'_2) \quad (5)$$

and simplified by ignoring the very small intercept term.

$$L = 0.051(v'^2_1 - v_1^2) + 0.051\frac{m_2}{m_1}v'^2_2 - 0.051\frac{m_2}{m_1}v_2^2 \quad (6)$$

Even considering the values of the coefficients, this data-driven model is almost identical to the physically true model (equation 4). On the other hand, the NN model is an appropriate model whose basis functions are sigmoid and linear ones in this case.

Figures 5c and d show the prediction accuracies of the NN and RLS regression models for large-noise data ($\sigma$ = 0.15). An equation for the RLS regression model is as follows.

$$L = 0.131 - 0.082\left(\frac{v_2}{m_1} - \frac{v'_2}{m_1}\right)m_2 - 0.015(v'^2_1 - v_1^2)v'_1 + 0.003\frac{m_2}{m_1}v'^3_2 \quad (7)$$

Although the RLS regression failed to construct a physically true model (equation 4) because of the too large noise, the prediction accuracy of the RLS regression model is comparable to that of the NN model.

In summary, we proposed a recursive-feature-selection-based symbolic (RLS) regression method, where the generation of nonlinear features and feature selection are recursively repeated and, finally, highly nonlinear features can be created automatically. By using this method, we demonstrated that two simple natural laws including highly nonlinear relationships can be data-drivenly formulated as interpretable mathematical expressions without any prior knowledge. The RLS regression model can give us valuable inputs and impetus for constructing new physical rules leading to novel scientific discoveries.

**Methods**

**RLS regression.** The recursive-LASSO-based symbolic (RLS) regression method can construct a data-driven nonlinear mathematical expression model. This method includes iteration of a nonlinear-features generation step and feature-selection step. For simplicity of explanation, we here assume only three original features $x = \{x_1, x_2, x_3\}$. In the nonlinear-features-generation step, nonlinear features $x_g$ are created by the four basic arithmetic operators (and the squaring operator) with the original explanatory variables $x$. Figure 2a is the list of $x_g$. In the feature-selection step, the $x_g$ is screened by a LASSO-like algorithm, where the evaluation function of LASSO[24] is modified as follows.

$$\|y - X_g\beta\|_2^2 + \lambda \sum_{m=1}^{M} A_m(\sigma)\|\beta_m\|_1 \quad (8)$$

$y$, $X_g$, $\beta$, and $\lambda$ are respectively an objective variable vector, a design matrix of nonlinear features $x_g$, a regression coefficients vector, and a regularization parameter. $M$ is the number of nonlinear features $x_g$, and $\sigma$ is the number of original features ($x_1, x_2, x_3$) in each nonlinear feature. For example, $\sigma$ of $x_3$, $\frac{x_2 x_3}{x_1}$, and $\frac{x_3 x_1^3}{x_2 - x_3}$ are respectively 1, 3, and 6. With the $\sigma$, the $A_m(\sigma)$ controls the magnitude of regularization for each nonlinear feature. Here, we just use a step function as follows.

$$A_m(\sigma) = \begin{cases} 10 & (\sigma > \sigma') \\ 1 & (\sigma \leq \sigma') \end{cases} \quad (9)$$

In this study, $\sigma'$ was set to the number of original features $x$ (i.e., $\sigma'$ was 4 for the small-particle's-falling case study and was 7 for the rigid-body-collision case study). The function $A_m(\sigma)$ prevents the selection of nonlinear features that are too complex (e.g., $\frac{x_3 x_1^3 - \frac{x_2^4}{x_1}}{x_2 - x_3 + x_1^3}$). Exploring a model searching space including features that are too complex is a waste of time because equations of natural laws are often simple as well as nonlinear. Nonlinear features $x'$, some of which can be a bit more complex than the original features $x$, are selected in the feature-selection step by minimizing equation (8). Here we assume that three terms $x' = \{x_2, x_1 x_2, x_2 + x_3\}$ are selected. Again, the RLS regression makes some nonlinear features by using both original features $x$ and the selected nonlinear features $x'$ with de-duplication (i.e., $x_2$). Figure 2b shows the nonlinear features generated in the second feature-selection step. Note that they are a bit more complex than the features generated in the first one (Fig. 2a). The RLS regression automatically searches for proper nonlinear features by recursively repeating this cycle. By using a 4-fold cross-validation error, a final model is selected from among models (with $x'$) created in each iteration. If the number of $x'$ is too large, it takes a long time to calculate the next nonlinear-features-generation step and feature-selection step. Therefore we provide an upper limit number $K$ of $x'$ in the LASSO-like algorithm. The parameters in the RLS regression algorithm are the same in both case studies (small particle falling and rigid-body collision). The upper limit number $K$ of $x'$ in the LASSO-like algorithm was set to 40. The threshold $\sigma'$ in $A_m(\sigma)$ is 2 + the number of **x**. For the $A_m(\sigma)$ setting, "penalty.factor" of the cv.glmnet function in the glmnet package in R programming language was used. For the 4-fold cross-validation, mean absolute error (MAE) was used.

This RLS regression algorithm has room for customization, and here we mention three examples. One would be to expand nonlinearity. In the nonlinear-features-generation step, we consider just four simple arithmetic operators and the square operator. However, we could include other operators, such as those for exponential and trigonometric functions, in this step so that RLS regression could make more complex features such as $x_1 e^{-\frac{x_3}{x_2}}$ and $\frac{x_3}{1 + x_1 \cos(x_2)}$. The second customization would be to sophisticate the $A_m(\sigma)$. In this study, we used a quite simple function, the step function defined as shown in equation (9). However, we can design an advanced $A_m(\sigma)$ for more effective model exploration. The third customization would be to employ other feature selection methods instead of the LASSO-like algorithm. For example, maximal information coefficient (MIC)[25] could be used in the feature selection part to improve the RLS regression modeling.

**Acknowledgements**

This work was supported by JST-PRESTO "Advanced Materials Informatics through Comprehensive Integration among Theoretical, Experimental, Computational and Data-Centric Sciences" (Grant No. JPMJPR17N4).

**Figures**

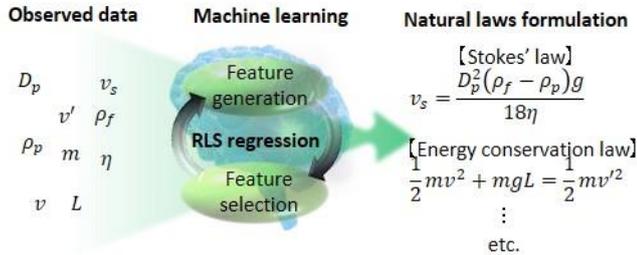

**Figure 1 | Overview of data-driven formulation using the RLS regression.** Without prior knowledge about natural science, machine learning can construct a data-driven model from observed big data. The proposed machine learning method, recursive-feature-selection-based symbolic (RLS) regression, repeats feature-generation and feature-selection steps recursively and constructs interpretable mathematical expression models of highly nonlinear phenomena.

**Figure 2 | Explanation of the RLS regression algorithm.** When we assume three original features $x = \{x_1, x_2, x_3\}$, **a.** the list of nonlinear features generated by the first feature-generation step using three original features. After that, feature selection, e.g., LASSO, is carried out concerning the 21 features. When assuming the three features $x' = \{x_2, x_1x_2, x_2+x_3\}$ are selected by the first feature-selection step, **b.** the list of features generated by the second feature-generation step using original features $x = \{x_1, x_2, x_3\}$ and selected features $x' = \{x_2, x_1x_2, x_2+x_3\}$ by the first feature-selection step. By repeating this cycle recursively, the RLS regression automatically searches for proper nonlinear features. Details of this algorithm and parameter settings are provided in the Methods section.

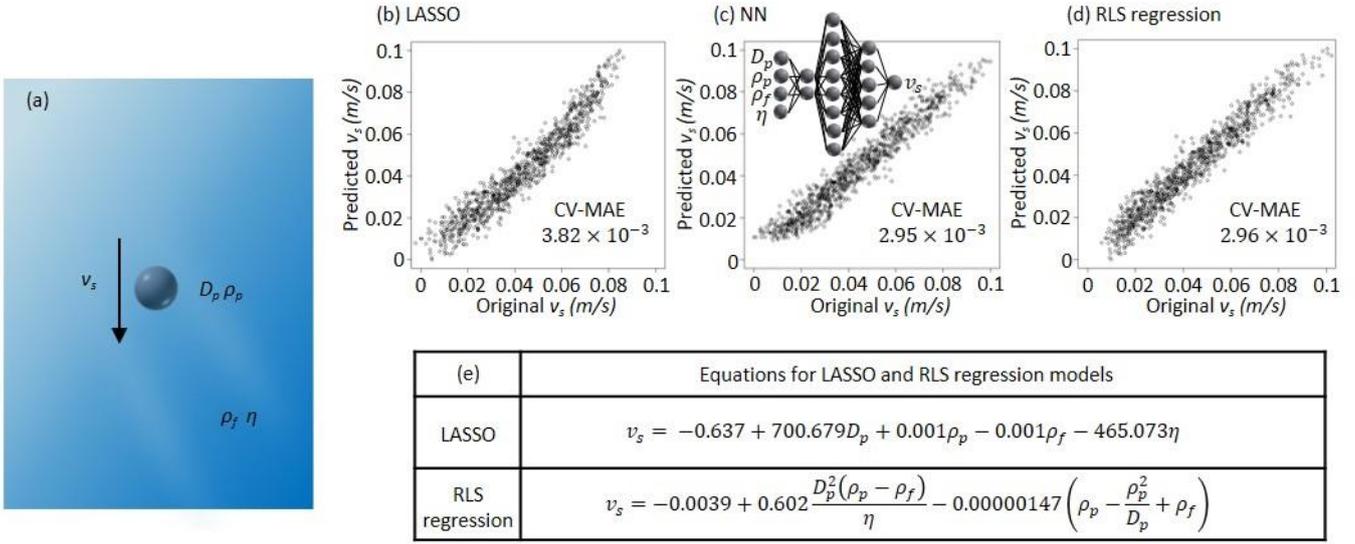

**Figure 3 | Data-driven formulation of small particles falling in fluid. a.** Illustration of falling particle with the particle radius $D_p$, particle density $\rho_p$, fluid density $\rho_f$, fluid viscosity $\eta$, and particle terminal velocity $v_s$. The prediction accuracy of the data-driven model $v_s = f(D_p, \rho_p, \rho_f, \eta)$ using **b.** LASSO, **c.** neural network NN (the model visualization is also shown), and **d.** RLS regression. The NN and RLS regression models have better prediction accuracy than the LASSO model. **e.** Equations for LASSO and RLS regression models. The RLS regression constructs the data-driven model including highly nonlinear features. In this case, Stokes' law was automatically constructed at the second term of the RLS regression model.

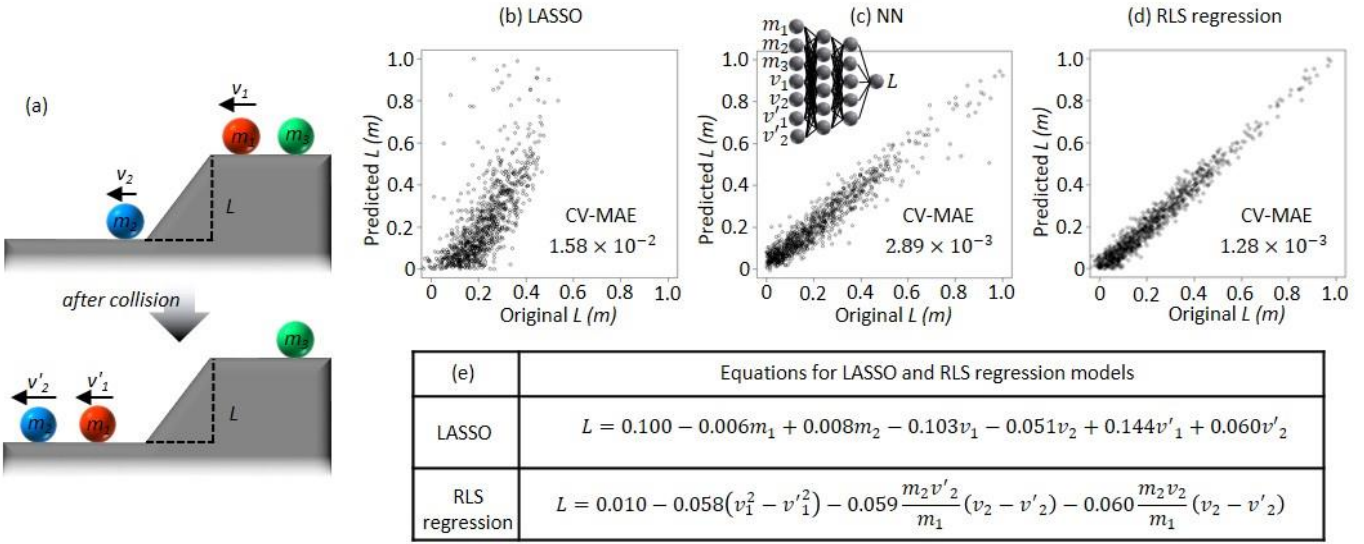

**Figure 4 | Data-driven formulation of rigid-body collision. a.** Illustration of rigid-body (ball) collision. Firstly, a red ball with mass $m_1$ is rolling on the hill at a velocity $v_1$. At the foot of a slope with height L, a blue ball with mass $m_2$ is rolling at a velocity $v_2$. After a few seconds, the red ball accelerates on the slope and collides with the blue ball. Finally, the velocities of the red and blue balls become $v_1'$ and $v_2'$, respectively. **b.** The prediction accuracy of the data-driven model $L = f(m_1, m_2, m_3, v_1, v_2, v'_1, v'_2)$ using **b.** LASSO, **c.** neural network NN (the model visualization is also shown), and **d.** RLS regression. The NN and RLS regression models have better prediction accuracy than the LASSO model. e. Equations of LASSO and RLS regression models. The RLS regression constructs the data-driven model including highly nonlinear features. In this case, the law of the conservation of energy was automatically constructed in the RLS regression model.

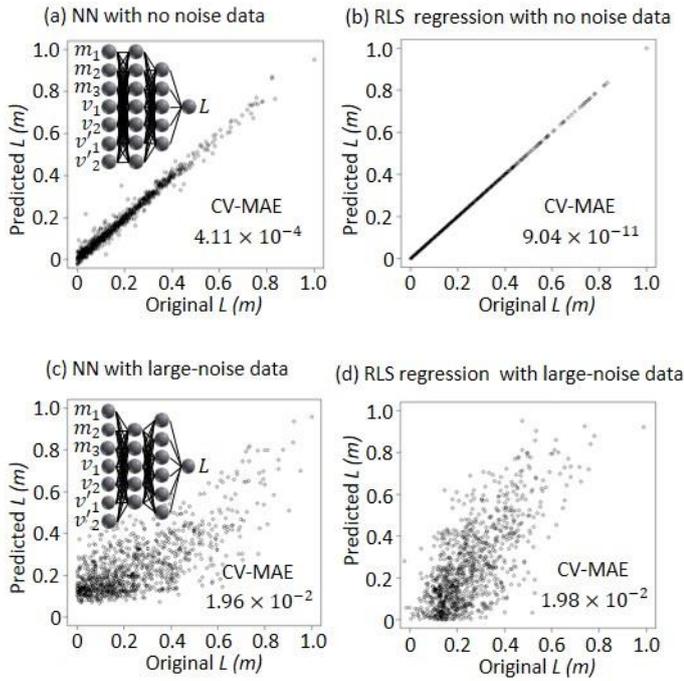

**Figure 5 | Investigation of noise robustness of the RLS regression model.** In the case study of the rigid-body-collision modeling, the noise robustness of the NN and RLS regression models was investigated. **a.** The prediction accuracy of the NN model for no-noise data ($\sigma = 0$). **b.** The prediction accuracy of the RLS regression model for no-noise data ($\sigma = 0$). The RLS regression model has a quite high prediction accuracy because it is a physically true model. **c.** The prediction accuracy of the NN model for large-noise data ($\sigma = 0.15$). **d.** The prediction accuracy of the RLS regression model for large-noise data ($\sigma = 0.15$). The prediction accuracy of the RLS regression model is comparable to that of the NN model.